\documentclass[aps,prl,reprint]{revtex4-1}

\pdfoutput=1 % for arXiv

\usepackage{graphicx}
\usepackage{bm}
\usepackage{dcolumn}
\usepackage{amssymb}
\usepackage{amsmath}
\usepackage{amsfonts}
\usepackage{color}
\usepackage{epsfig}
\usepackage{hyperref}
\usepackage{verbatim}

\def \No{N_{\mathrm{o}}}
\def \Ni{N_{\mathrm{i}}}
\def \Ko{K_{\mathrm{o}}}
\def \Ki{K_{\mathrm{i}}}
\def \Koi{K_{\mathrm{oi}}}
\def \Di{\Delta_{\mathrm{i}}}
\def \da{^{\dagger}}

\begin{document}

\title{ Transmission phase lapses through a quantum dot in a strong magnetic 
field }

\author{Yehuda Dinaii}
\affiliation{Department of Condensed Matter Physics, The Weizmann Institute of
             Science, Rehovot 76100, Israel}
\author{Yuval Gefen}
\affiliation{Department of Condensed Matter Physics, The Weizmann Institute of
             Science,  Rehovot 76100, Israel}
\author{Bernd Rosenow}
\affiliation{Institut f\"ur Theoretische Physik, Universit\"at Leipzig, D-04103,
             Leipzig, Germany}

\begin{abstract}

The phase of the transmission amplitude through a mesoscopic system contains 
information about the system's quantum mechanical state and excitations 
thereof. In the absence of an external magnetic field, abrupt phase lapses occur 
between transmission resonances of quantum dots and can be related to the 
signs of tunneling matrix elements. They are smeared at finite temperatures.  
By contrast, we show here that in the presence of a strong magnetic field, 
phase lapses represent a genuine interaction effect and may occur also on 
resonance.  For some realistic parameter range these phase lapses are robust 
against finite temperature broadening.

\end{abstract}

% \pacs{73.43.-f, 73.43.Cd, 73.20.-r, 71.23.-k} %

\maketitle

%====== Introduction

The evolution of the transmission phase of an electron traversing a small 
electron droplet, i.e.~a quantum dot (QD), has been the subject of intense 
research in the past two decades~\cite{Yacoby_1995, Schuster_1997, Ji_2000,
Avinun-Kalish_2005, Taniguchi_1999, Yeyati_2000, Silva_2002, Meden_2006, 
Golosov_2006, *Golosov_2007, Karrasch_2007, Goldstein_2007, *Goldstein_2009, 
*Goldstein_2010, Hackenbroich_1997, Baltin_1999c, Silvestrov_2000, 
*Silvestrov_2001, Berkovits_2005, Goldstein_2009, Goldstein_2010}.  In the 
absence of a magnetic field the transmission phase exhibits a continuous and 
monotonic evolution as one sweeps through a transmission resonance. More 
interestingly, it jumps abruptly between transmission peaks. These so-called 
\emph{phase lapses} can be explained in the framework of non-interacting 
electrons, if one considers the sign and magnitude of the hopping matrix 
elements connecting the QD to its two leads~\cite{Silva_2002, Golosov_2006, 
Golosov_2007, Karrasch_2007}. The ubiquity of phase lapses may invoke the 
presence of intra-dot interactions, and could be related to the mechanism of 
\emph{population switching}: an abrupt ``swap'' of two level occupations as 
the gate voltage is varied~\cite{Hackenbroich_1997, Baltin_1999c, 
Silvestrov_2000, Silvestrov_2001, Berkovits_2005, Konig_2005, *Sindel_2005, 
Goldstein_2009, Goldstein_2010}. We also note theories that invoked 
correlations due to the chaotic nature of the QD~\cite{Molina_2012, 
*Molina_2013, Baltin_1999a, *Baltin_2000}.

% \cite{*[{text}] [{text}] ref}

In the presence of a strong magnetic field~\footnote{The effect of weak
magnetic fields has been noted in~\cite{Silva_2002} and in~\cite{Molina_2012, 
*Molina_2013}}, specifically in the integer quantum Hall (QH) regime, the 
aforementioned picture is likely to change.  This has to do with the chiral 
motion of electrons along equi-potential contours inside the QD, forming one 
dimensional edge states~\cite{Halperin_1982, MacDonald_1990, Wen_1990, 
Wen_1991}. In this regime electrons cannot backscatter off impurities (unless 
a counter-propagating edge is nearby).  Moreover, the magnetic-field-acquired 
phase of the wave functions cannot be gauged out, rendering the tunneling 
matrix elements complex.
Do phase lapses occur under such circumstances too?

\begin{figure}[tbp]
\begin{center}
% \vspace{-0.5cm}
\includegraphics[width=0.45\textwidth]{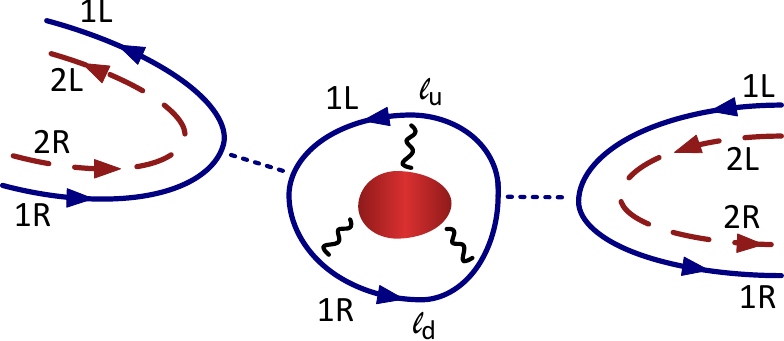}
\vspace{-0.3cm}
\caption[] { \label{fig:setup}
A gate-defined quantum dot (QD) operating in the quantum Hall regime (filling 
factor $\nu = 2$). The QD consists of two parts. The outer channels (denoted 
by 1R and 1L, blue solid lines) form a ring shaped region. Tunneling between 
this region and the associated channels outside the QD is denoted by  
dotted blue lines. The inner channels (denoted by 2R and 2L, red dashed lines) 
define an isolated puddle (or state) tunnel-coupled to  leads (the coupling is 
not shown in the figure).  Wiggly lines represent the electrostatic 
interaction between the localized puddle and the outer edge mode of the QD. 
The parameter $\ell_\mathrm{u}$ ($\ell_\mathrm{d}$) denotes the length of the 
upper (lower) arm of the outer channel in the QD.  The transmission amplitude 
of electrons traveling along channel 1R towards the dot is measured by 
embedding it in one arm of a Mach-Zehnder interferometer (not shown).
}
\end{center}
\end{figure}

We present here a study of a QD operating in the QH regime with filling factor 
$\nu = 2$, where the Hall bar supports two co-propagating edge 
channels~\cite{Halperin_1982, *MacDonald_1990, *Wen_1990, *Wen_1991, 
Prange_1990, *Das_Sarma_1997}. One outer channel (1R, cf.  
Fig.~\ref{fig:setup}) is set to be part of the arm of a Mach-Zehnder 
interferometer (MZI).  This facilitates the measurement of the complex 
transmission amplitude through the QD~\cite{Weisz_2012}. Here we find that (i) 
phase lapses may occur also in this regime of a strong magnetic field, but 
that the underlying physics is utterly different from the zero field case.  
Importantly, these phase lapses represent a genuine many-body effect, 
resulting from the interaction between the inner and outer edge channels
(the inner edge channel may also be represented by an orbital level or a 
compressible puddle). (ii) In the standard case, zero transmission and phase 
lapses are due to the coherent addition of two or more transmission amplitudes 
through the quantum dot. In contradistinction, in the strong magnetic field 
case phase lapses are due to true dephasing as an internal degree of freedom 
fluctuates inside the quantum dot.  (iii) For  zero magnetic field phase 
lapses acquire a width $\sim \mathsf{T}^2$  at a finite temperature $\mathsf{
T}$ \cite{Silva_2002}. By contrast we find that for a realistic, 
experimentally relevant parameter  range,  strong magnetic field phase lapses  
are robust against broadening at finite temperatures.

%======  Body

% === Figure 1 : setup >>>

Two gate controlled constrictions in the Hall bar form a QD (cf.  
Fig.~\ref{fig:setup}). In the QH regime with $\nu = 2$ the electrons move 
inside the QD along two chiral edge modes. We focus on the transmission of the 
outer channel.  Assuming that the  magnitude of charge fluctuations on the 
inner mode (the localized puddle in Fig.~\ref{fig:setup})  do not exceed an 
electron charge, it is reasonable to treat it as a localized level which may 
be either occupied or empty.  The spatial structure of the outer edge channel 
of the QD is important, and in what follows will be taken into account.
The gates at the left and right sides of the dot control the corresponding 
tunneling amplitudes.  Tunneling between the two edge channels is suppressed 
as these correspond to oppositely spin polarized modes.  The respective 
couplings of the outer channel and the inner puddle to the external edge modes 
define two time scales, namely the typical times for charge fluctuations in 
the corresponding region. It will be assumed that during the passage of one 
electron through the outer region, the localized level's occupation remains 
unchanged, i.e. each passing electron through the outer region senses the 
localized level as a non-dynamical environment~\cite{Stern_1990}.

Our aim is to calculate the transmission amplitude through the QD.  We first 
consider the zero temperature quantum regime, and later will generalize our 
discussion to finite temperatures.  The effect of the localized level is to 
provide an electron passing through the outer region of the QD with an extra 
phase, if this level is occupied~\cite{Rosenow_2012}.  Specifically, an 
electron occupying the localized level induces a change in the density of 
electrons at channels 1R and 1L.  Employing the Thomas-Fermi approximation, 
this change is $\delta \rho_{\mathrm{R/L}}(x) = - e 
V_{\mathrm{R}/\mathrm{L}}(x) / {2 \pi \hbar v}$, where 
$V_{\mathrm{R}/\mathrm{L}}(x)$ is the potential induced in channel 1R(1L) by 
the electron occupying the localized level, whose charge is $e < 0$; $x$ is 
the spatial coordinate along the corresponding channel, and $v$ is the 
velocity of electrons along the channel. When the localized level is empty, an 
electron at the Fermi level $\varepsilon_{\mathrm{F}}$ acquires a phase 
$\varepsilon_{\mathrm{F}} \Delta x / \hbar v$ while traversing a distance 
$\Delta x$. In the presence of the potential $V_{\mathrm{R/L}} (x)$, i.e.  
when the localized level is occupied, the chemical potential changes locally 
by $- e V_\mathrm{R/L} (x)$. This, in turn, induces an extra phase equal to $- 
e  \int_0^{\Delta x} \mathrm{d}x \, V_\mathrm{R/L} / \hbar v = 
2 \pi \int_0^{\Delta x} \mathrm{d}x \, \delta \rho_{\mathrm{R/L}}(x) \equiv
\theta_{\mathrm{R}} (\theta_{\mathrm{L}})$, where $\theta_{\mathrm{R}} + 
\theta_{\mathrm{L}} = 2 \pi$. The last equality reflects the fact that the 
total screening charge is $e$. For symmetric screening between channels 1R and 
1L, $\theta_{\mathrm{R}} = \theta_{\mathrm{L}} = \pi$. Similarly, we define 
the screening phase $\theta$, which denotes the extra phase accumulated by an 
electron while winding once along channels 1R and 1L inside the QD. It turns 
out that the results of our calculation can be formulated using only the 
screening phases $\theta$ and $\theta_\mathrm{R}$.

The spatial dependence of $\delta \rho_{\mathrm{R/L}} (x)$ and the ensuing 
screening phase is important for the analysis of the transmission amplitude.  
We note that part or all of the screening takes place inside the QD. Then  
multiple winding trajectories imply multiple accumulation of the screening 
phase  $\theta$.  Clearly, $0 \le \theta \le 2\pi$, where $\theta / 2\pi$ is 
the fraction of electron charge screened inside the QD.  Similarly, $0 \le 
\theta_{\mathrm{R}} \le 2\pi$, where $\theta_{\mathrm{R}} / 2\pi$ is the 
fraction of electron charge screened along channel 1R. Below we assume that 
screening does not take place along channels 2R and 2L (generalization beyond 
this assumption is straightforward).

In order to measure the transmission amplitude through the QD, the latter is 
embedded in one arm of a MZI (``upper''). The wave packet of an electron 
injected into the MZI is split into two upon arriving at its first junction.  
The lower partial wave, $\left| \mathrm{d} \right\rangle$, goes directly 
towards the second junction and interferes with the part of the upper partial 
wave that is transmitted through the QD, $\left| \mathrm{u} \right\rangle$.  
The current through the MZI as measured at one of its drains is proportional 
to the probability of an electron to arrive at that drain. Thus, the current 
is a function of the transmission phase through the QD.

The scattering matrix of the QD
depends on the initial state of the isolated subsystem consisting of the 
localized level and the tunnel-coupled lead(s) (red dashed lines and puddle in 
Fig.~\ref{fig:setup}).  Formally, this state is a Slater determinant built of 
the eigenstates of that subsystem.  Here, we do not include the interaction 
between the localized state and the outer edge mode of the QD
since it does not change our picture in a qualitative manner.  However, it is 
possible to show~\footnote{See Supplemental Material for a discussion about 
  the initial wave function of the subsystem consisting of the localized state 
  and the tunnel-coupled lead, the mean occupation $n$, the density matrix 
  $\hat{\rho}$, the projection operator $\hat{D}$, and the charging energy 
model.} that this subsystem can be treated as a two-state system, whose wave 
function is $\sqrt{1-n} \left| 0 \right\rangle + \sqrt{n} \left| 1 
\right\rangle$. Here $\left| \sigma \right\rangle$ is a basis state vector 
corresponding to an empty ($\sigma = 0$) or occupied ($\sigma = 1$) localized 
level; an unimportant relative phase factor is omitted.  Due to the fermionic 
statistics of the electrons, the probability of the localized level to be 
occupied, $n$, equals its \emph{mean occupation}.  The calculation of $n$ is 
elementary~\footnotemark[2], \cite{Mahan_3ed}.  The result is
\begin{equation}
n = \frac{1}{\pi}\left[\arctan\left(\frac{\mu-\epsilon_{0}}{\Gamma}\right)
+\frac{\pi}{2}\right] \,,
\label{eq:red_dot_occupation_probability_T=0_result}
\end{equation}
where $\mu$ is the chemical potential of the system, $\epsilon_0$ the 
eigenenergy of the localized state, and $\Gamma$ its width due to the 
tunnel-coupled leads.

We calculate~\footnotemark[2] the transmission amplitude through the QD 
employing scattering matrices and taking into account properly the extra 
phases $\theta$ and $\theta_{\mathrm{R}}$. If the localized level is occupied, 
the transmission amplitude through the QD for an electron traveling along the 
channel 1R is
\begin{equation}
  t_{\mathrm{QD}}(\epsilon,\theta,\theta_{\mathrm{R}}) = \frac{ \gamma e^{i( 
    \epsilon \ell_\mathrm{d}/\ell + \theta_{\mathrm{R}})} } {1 - e^{i(\epsilon 
    - \theta)} + \gamma} \,.
  \label{eq:QD_transmission_amplitude}
\end{equation}
Here $\ell = \ell_\mathrm{d} + \ell_\mathrm{u}$ is the circumference of the 
outer channel inside the QD, which is the sum of the lower ($\ell_\mathrm{d}$) 
and upper ($\ell_\mathrm{u}$) lengths (see Fig.~\ref{fig:setup}). The 
dimensionless parameter $\epsilon \equiv 2 \pi \alpha V_g / \Delta$ , shifting 
the outer region energy levels, is proportional to the gate voltage $V_g$ with 
lever arm $\alpha > 0$. $\Delta = 2 \pi \hbar v / \ell$
is the level spacing in the \emph{bare}  outer region, namely in the absence 
of the inner puddle.  The dimensionless parameter $\gamma$ reflects the width 
of the levels of the  outer part of the QD in the absence of the inner puddle.  
The phase $\theta_{\mathrm{R}}$ accounts for the fact that part of the 
screening takes place  on channel 1R outside the QD.  The ensuing phase of the 
expression in~(\ref{eq:QD_transmission_amplitude}) is the added contributions 
accumulated inside and outside the QD. In a generic case (beside the cases 
$\ell_{\mathrm{d}}/\ell = 0, 1$ which are unfeasible), and in  a situation 
where  the phases $\theta$ and $\theta_{\mathrm{R}}$ do not vary with energy, 
the transmission amplitude described by 
Eq.~(\ref{eq:QD_transmission_amplitude}) does not have phase lapses. 

% === Figure 2 : transmission phase and coherence >>>
\begin{figure}[tbp]
\begin{center}
\vspace{0cm}
\includegraphics[width=0.49\textwidth]{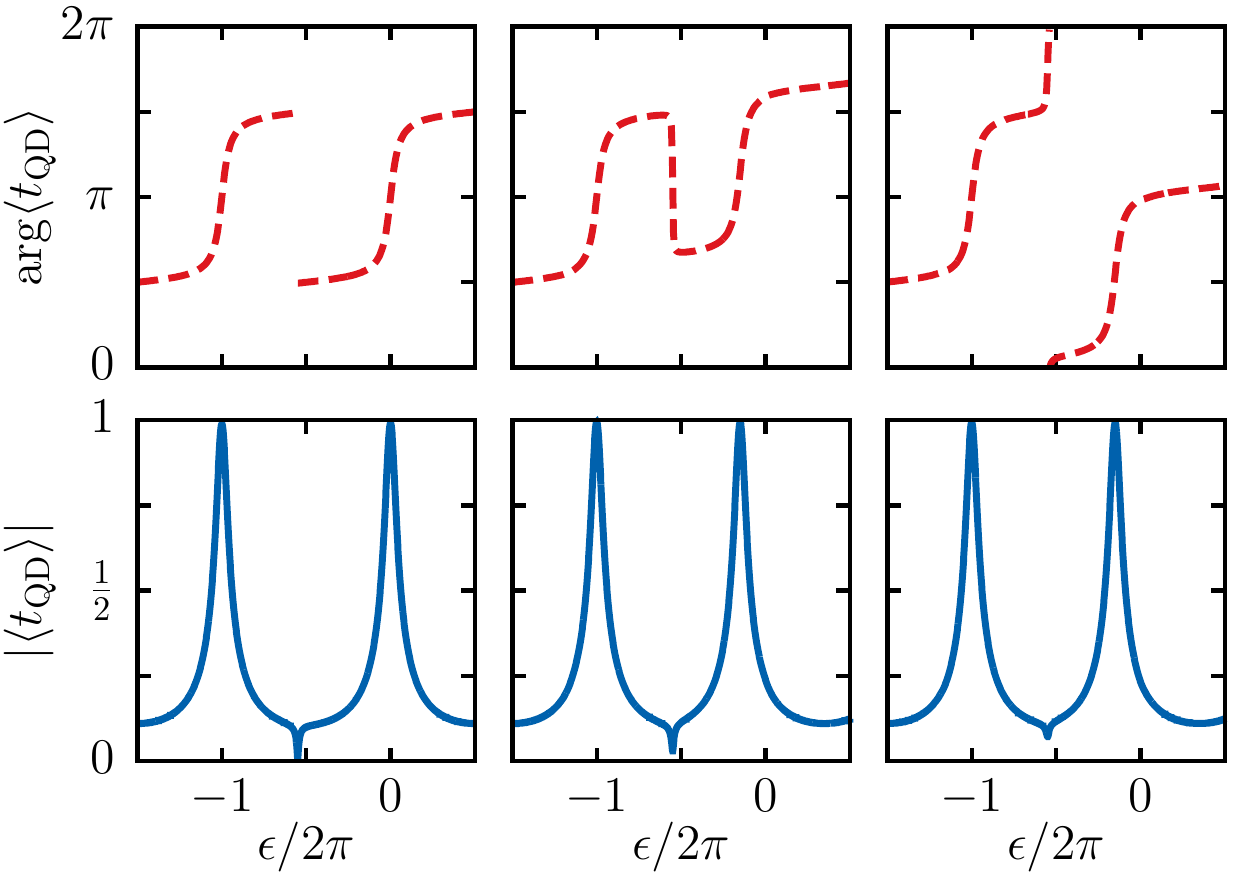}
\vspace{-0.5cm}
\caption[] {\label{fig:vis_phase_closed}
Transmission phase through the QD (red dashed lines, top) and magnitude of 
coherent oscillations in the MZI (blue solid lines, bottom) in the strong 
coupling regime for the sequence $(\No, \Ni) \rightarrow (\No + 1, \Ni) 
\rightarrow (\No, \Ni + 1) \rightarrow  (\No + 1, \Ni + 1)$  as a function of 
the dimensionless energy parameter $\epsilon$ for symmetric 
($\theta_{\mathrm{R}} = \pi$, left) and slightly asymmetric 
($\theta_{\mathrm{R}} = 1.3 \pi$ (center) and $\theta_{\mathrm{R}} = 0.7 \pi$ 
(right)) setups. The left (center, right) plot depicts a sharp (smeared) phase 
lapse accompanied by a full (partial) suppression of the coherent oscillations 
through the MZI. The energy level of the outer region and the occupancy of the 
inner region (cf.  Eq.~(\ref{eq:red_dot_occupation_probability_T=0_result})) 
are controlled by a common gate voltage $V_g$ with lever arms $\alpha$ and 
$\beta$ through the relations $\epsilon = 2 \pi \alpha V_g / \Delta$ and $(\mu 
- \epsilon_0) / \Gamma = (\beta V_g + c) / \Gamma$, respectively, where 
$\Delta$ is the level spacing in the outer region, $\Gamma$ is the level width 
in the inner region, $(\beta / \alpha) \cdot (\Delta / 2 \pi \Gamma) = 20$ and 
$c / \Gamma = 22 \pi$.
Here, using dimensionless parameters, the level spacing of the outer edge (in 
the absence of the inner edge) is $2\pi$. We consider the symmetric case 
$\ell_{\mathrm{d}}/\ell = 1/2$, and weak coupling to the leads, $\gamma = 
1/4$. The screening phase $\theta = 2\pi$ (left) or $\theta = 
1.7\pi$ (center and right) imply full and almost full screening inside the
QD, respectively.
} % caption
\end{center}
\vspace{0.0cm}
\end{figure}

The transmission probability through the MZI is obtained by employing a pure 
state density matrix~\footnotemark[2]. This yields
\begin{multline}
  T = \mathrm{Tr}\left(\hat{\rho}\hat{D}\right) = \frac{1}{4} + \frac{1}{4} 
  \left[ (1-n) \left| t_{\mathrm{QD}}(0) \right|^{2} + n\left| 
    t_{\mathrm{QD}}(1) \right|^{2} \right] \\
  + \frac{1}{2} \Re \left[ e^{-i\phi} \left[ (1-n) t_{\mathrm{QD}}(0) + n 
    t_{\mathrm{QD}}(1) \right] \right] \\
      = \frac{1}{4} + \frac{1}{4}\left\langle \left| t_{\mathrm{QD}} 
      \right|^{2} \right \rangle + \frac{1}{2} \Re \left[ e^{-i\phi} 
        \left\langle t_{\mathrm{QD}} \right\rangle \right] \,.
  \label{eq:MZI_transmission_probability}
\end{multline}
Here $\hat{\rho}$ is a density matrix constructed from the wave function of the 
whole system. It corresponds to the interfering electron being either 
scattered by the QD or transmitted through the lower MZI arm.
The operator $\hat{D}$ is defined by $\left(\left\langle 
\sigma\right|\otimes\left\langle s \right|\right)\hat{D}\left(\left| s' 
\right\rangle \otimes\left|\sigma'\right\rangle 
\right)=\delta_{\sigma\sigma'}/2$
  for all combinations of $s, s' = \mathrm{u}, \mathrm{d}$.
The  operator $\hat{D}$ has two functionalities, namely it selects only the 
part of the wave function that arrives at the measured drain, and taking the 
trace over $\hat{D}$ integrates out the environmental degrees of
freedom~\footnotemark[2].  The phase $\phi = 2 \pi \Phi / \Phi_0$, where 
$\Phi$ is the magnetic flux enclosed by the MZI arms, $\Phi_0 = 2 \pi \hbar c 
/ |e|$ the magnetic flux quantum, $e$ the charge of an electron and $c$ the 
speed of light. The transmission amplitudes $t_{\mathrm{QD}}(1)$ and 
$t_{\mathrm{QD}}(0)$ are abbreviations for 
$t_{\mathrm{QD}}(\epsilon,\theta,\theta_{\mathrm{R}})$ and 
$t_{\mathrm{QD}}(\epsilon,0,0)$, respectively (cf.  
Eq.~(\ref{eq:QD_transmission_amplitude})). It should be emphasized that our 
calculation is valid in the regime where the time interval between two 
consecutive transmitted electrons is sufficient for the inner puddle to relax 
to its ground state. In the third line of 
Eq.~(\ref{eq:MZI_transmission_probability}) and henceforth angular brackets 
$\langle \ldots \rangle$ denote the average value of the quantity inside the 
brackets, calculated with respect to the probability distribution function
\begin{equation}
  P(\tilde{\theta}, \tilde{\theta}_{\mathrm{R}})
  = \begin{cases}
    n   & \mathrm{for\; the\; phases\; to\; be\;(\theta, 
      \theta_{\mathrm{R}})}\,,\\
    1-n & \mathrm{for\; the\; phases\; to\; be\;(0, 0)}\,.
  \end{cases}
\label{probability_distribution_function}
\end{equation}
The parameters $\theta$ and $\theta_{\mathrm{R}}$ are defined above, and 
corresponding random variables are denoted by $\tilde{\theta}$ and 
$\tilde{\theta}_{\mathrm{R}}$. Thus the last equality in 
Eq.~(\ref{eq:MZI_transmission_probability}) shows that the presence of the 
localized state turns the transmission amplitude of the QD into a random 
quantity, whose probability distribution function is determined by $n$ (cf.  
Ref~\cite{Stern_1990}).

The two quantities of interest are the transmission phase through the QD,
$\arg \left\langle t_{\mathrm{QD}} \right\rangle$, and the magnitude of the 
coherent oscillations of the current through the MZI, $\left| \left\langle 
t_{\mathrm{QD}} \right\rangle \right|$.  From 
Eq.~(\ref{eq:MZI_transmission_probability}) we find
\begin{subequations}\label{eq:grp}
  \begin{align}
  \arg \left \langle t_{\mathrm{QD}} \right \rangle & = \arg \left[
   t_{\mathrm{QD}}(\epsilon, 0,0) \left \langle \zeta \right \rangle \right] 
   \,,
  \label{QD_transmission_phase_result}\\
  \left| \left \langle t_{\mathrm{QD}} \right \rangle \right| & = \left| 
  t_{\mathrm{QD}}(\epsilon, 0,0) \right| \left| \left \langle \zeta \right 
  \rangle \right| \,,
  \label{eq:vis}
  \end{align}
\end{subequations}
% \vspace{0.8cm}
%
\begin{comment}
\begin{equation}
  \arg \left \langle t_{\mathrm{QD}} \right \rangle = \arg \left[
   t_{\mathrm{QD}}(\epsilon, 0,0) \left \langle \zeta \right \rangle \right]
  \label{QD_transmission_phase_result}
\end{equation}
and 
\begin{equation}
  \left| \left \langle t_{\mathrm{QD}} \right \rangle \right| = \left| 
  t_{\mathrm{QD}}(\epsilon, 0,0) \right| \left| \left \langle \zeta \right 
  \rangle \right| \,,
\label{eq:vis}
\end{equation}
\end{comment}
%
where
\begin{equation}
  \zeta ( \epsilon, \tilde{\theta}, \tilde{\theta}_{\mathrm{R}}  ) = \frac{1 + 
    \gamma - e^{i \epsilon}}{1 + \gamma - e^{i(\epsilon - \tilde{
  \theta})}} e^{i \tilde{\theta}_{\mathrm{R}}}\,.
\label{eq:zeta_and_a}
\end{equation}
Here averages are calculated with respect to the probability 
distribution~(\ref{probability_distribution_function}), e.g.
\begin{equation}
  \left\langle \zeta   \right\rangle = 1 - n + n \zeta ( \epsilon, \theta, 
  \theta_{\mathrm{R}}  )  \, .
  \label{eq:zeta_average}
\end{equation}
Clearly, the presence of the inner puddle induces a change in the transmission 
phase such that $\arg \left[ t_{\mathrm{QD}}(\epsilon, 0,0) \right] 
\rightarrow \arg \left[ t_{\mathrm{QD}}(\epsilon, 0,0) \left \langle \zeta
\right \rangle \right]$. The resulting phase is the sum of the transmission 
phase through the ``bare'' outer region and $\arg( \left\langle \zeta 
\right\rangle)$.  Phase lapses can occur only due to the phase evolution of 
$\left \langle \zeta \right \rangle$, since $\arg \left[
t_{\mathrm{QD}}(\epsilon, 0,0) \right]$ by itself evolves continuously.  
Moreover, it is evident that any  interesting physics that may be hidden in 
$\left \langle \zeta \right \rangle$ will generically be more pronounced if it 
happens to occur in between resonances of the ``bare'' outer region --- there 
the phase of $t_{\mathrm{QD}}(\epsilon, 0,0)$ is practically constant.

A phase lapse occurs if $\left \langle \zeta \right \rangle$ vanishes at a 
certain $V_g$.  Eq.~(\ref{eq:vis}) shows that this abrupt jump in the phase is 
accompanied by complete suppression of the coherent oscillations in the MZI.  
Solution of the complex equation  $\left \langle \zeta \right \rangle = 0$ 
requires both $(1-n)/n = \left| \zeta \left( \epsilon, \theta, 
\theta_{\mathrm{R}}  \right) \right|$ and fine tuning of the phases $\theta$ 
and $\theta_{\mathrm{R}}$ such that $\arg \left[ \zeta \left(
\epsilon, \theta, \theta_{\mathrm{R}} \right) \right] = \pi$. These phases are 
determined by the geometry of the setup (the location of the localized level, 
the symmetry of the QD, etc ...), which fixes the way screening is divided in 
the system. The geometry can be controlled by tuning the gates that define the 
QD. Fig.~\ref{fig:vis_phase_closed} depicts the emergence of phase lapses in 
the transmission amplitude through the QD and the accompanying dephasing of 
the MZI.

This very general picture outlined above can be put to work employing
parameters that reflect the sample's specific electrostatic features. 
These parameters determine the effect of the gate voltage on the 
inner and outer parts of the QD, and hence the evolution of the transmission 
phase. Specifically, we employ a charging energy model~\footnotemark[2], which 
leads to a stability diagram of the charge distribution between the inner and 
outer parts of the QD, with charges  $\Ni$ and $\No$, respectively. 

In order to extract a physical picture out of this many-parameter
problem, we focus on two important limits, namely that of a
strong (``S'') and a weak (``W'') coupling between the two parts
of the QD. We examine each of these limits in view of two interesting scenarios
that may occur vis-\`a-vis the change in occupancy of the two parts
of the dot as a common gate voltage is varied~\footnotemark[2]. These 
scenarios are
(a)  $(\No, \Ni) \rightarrow (\No ,\Ni +1)$, and  (b)   $(\No ,\Ni) 
\rightarrow (\No - 1, \Ni + 1)$.

We begin with the strong interaction case (S), which implies 
$\theta\simeq2\pi$. In S(a) the outer part is positioned in a valley between 
resonances, while the inner part is tuned to be near a resonance peak and 
eventually crosses this peak as a function of gate voltage. 
Under these conditions a phase lapse occurs if 
$\theta_{\mathrm{R}}\simeq0,2\pi$~\footnotemark[2].
In S(b) there is a population switching (see e.g.~\cite{Baer_2013}). This 
means that both parts of the QD change their occupation by $\pm1$. If 
$\theta_{\mathrm{R}} \simeq \pi$, as appears to be achieved quite naturally in 
experiments~\cite{Weisz_2012}, then this scenario leads to a phase 
lapse~\footnotemark[2].

We turn now to the weak coupling regime, where $\theta \simeq 0$.  Scenario 
W(a) may occur when the outer channel is not too close to a resonance,
so that its occupation is not affected by a change in the occupation
of the inner puddle. Then there is no discontinuity
(yet possibly a sharp signature) in the transmission phase.
Scenario W(b), which implies a population switching, can occur only if the 
outer channel is close to a resonance. Then (in a generic case) a phase lapse 
occurs if $\theta_{\mathrm{R}}\simeq0,2\pi$, where  $\theta_{\mathrm{R}} 
\simeq 0$ is more likely in a weak coupling scenario. 

\emph{Finite temperatures ---} Our analysis so far pertains to the strictly  
zero temperature limit. Two modifications need to be introduced at finite 
temperatures: (i) The initial state of the subsystem composed of the localized 
level and the tunnel-coupled lead(s) must be described by a mixed density 
matrix (rather than a wave function).  This is easily handled as the operator 
$\hat {D}$  is diagonal in the localized level coordinate $\sigma$.  This 
implies that only the corresponding diagonal elements of $\hat \rho$ are of 
importance for the calculation of the transmission probability through the 
MZI. (ii) The electronic beam traveling along the arms of the MZI has finite 
width in energy. This means that  the entire interference pattern is a 
juxtaposition of many monochromatic partial beams; each such partial beam 
travels both in the upper and lower arm of the MZI. Summing over all 
contributions will naturally lead to thermal smearing and reduction  of the 
interference signal. This, however, is not our main focus here. We note that 
in scenario S(b) above, each such partial  interference would be shifted by a 
phase $ \pi$ due to the entry/exit of an electron to the localized level, and 
will be consequently fully dephased. That would mean that abrupt phase lapse 
accompanied by full dephasing will take place at finite temperature as well.  
This phase lapse and  dephasing will  take place on the background of an 
interference contrast which decreases with temperature.  We note that the 
physics is less simple with the other  scenarios outlined above. Charge 
fluctuations on the localized level will affect electron trajectories with 
different winding numbers differently. That would imply, in turn, that the 
efficiency of dephasing will vary with energy, leading to temperature 
dependent smearing of the phase lapses.

%====== Summary

To conclude, we have studied the transmission amplitude through a QD operating 
in the QH regime and have found that  it displays phase lapses due to 
interactions between different spin populations inside the dot.  Specifically, 
phase lapses occur  in the presence of quantum or thermal fluctuations, and 
are related to full dephasing of the electrons.  We have developed a formalism 
which allows to take into account the influence of both types of fluctuations  
in a unified way, and have identified the experimentally relevant regime of a 
strongly interacting and spatially symmetric
setup, where phase lapses are expected to occur due to population switching  
in the valley between transmission resonances. These phase lapses are not 
thermally broadened, in contrast to the zero magnetic field case.

We are grateful to E. Weisz and M. Heiblum for useful discussions, and to I. Chernii, I. Levkivskyi, and 
E. Sukhorukov for discussing with us their unpublished work. Financial 
support by the German-Israel Foundation (GIF), the Minerva Foundation,  the 
Israel Science Foundation, and the Federal Ministry of Education and Research (BMBF) is acknowledged.

\bibliography{ref}

\onecolumngrid

\pagebreak

\appendix

\section*{Supplemental Material}

\twocolumngrid

\setcounter{enumi}{1}
\setcounter{equation}{0}
\setcounter{figure}{0}
\renewcommand{\theequation}{\Roman{enumi}.\arabic{equation}}
\renewcommand{\thefigure}{\Roman{enumi}.\arabic{figure}}

% {\center \textbf{Transmission amplitude through the QD} }
\subsection*{Transmission amplitude through the QD }

We calculate the transmission amplitude through the QD employing scattering 
matrices and taking into account properly the extra screening phases $\theta$ 
and $\theta_{\mathrm{R}}$. If the localized level is occupied, the 
transmission amplitude through the QD of an electron traveling along channel 
1R is
\begin{equation}
  t_{\mathrm{QD}}(\epsilon,\theta,\theta_{\mathrm{R}}) = \frac{ t_{\mathrm{A}} 
  t_{\mathrm{B}} e^{i( \epsilon   \ell_\mathrm{d}/\ell + \theta_{\mathrm{R}})} 
} { 
      1 - r_{\mathrm{B}} r_{\mathrm{A}}' e^{i(\epsilon - \theta)} } \,.
  \label{eq:QD_transmission_amplitude_sup}
\end{equation}
Here $\ell = \ell_\mathrm{d} + \ell_\mathrm{u}$ is the circumference of the 
outer channel of the QD, which is the sum of the lower ($\ell_\mathrm{d}$) and 
upper ($\ell_\mathrm{u}$) lengths (see Fig.~1 in the main text). The 
dimensionless parameter $\epsilon \equiv \mu \ell / \hbar v$, where $\mu$ is 
the chemical potential of the system, and $v$ is the velocity of the 
electrons. To mimic the effect of a gate voltage $V_g$, capable of shifting 
the outer region energy levels, we take $\epsilon = 2 \pi \alpha V_g / \Delta$ 
with $\alpha > 0$ .The parameters $t_{\mathrm{A}}$, $t_{\mathrm{B}}$, 
$r_{\mathrm{A}}'$ and $r_{\mathrm{B}}$ are  elements of the scattering 
matrices associated with the left (A) and right (B) tunneling bridges that 
define the QD; here $t$ ($r$) refers to a transmission (reflection) amplitude 
of an electron propagating from left to right, and a prime denotes the 
opposite direction.  For example, $r_{\mathrm{B}}$ is the reflection amplitude 
of an electron, traveling along channel 1R inside the QD and impinging on the 
right junction from the left, to be reflected back to channel 1L inside the QD 
(Fig.~1 in the main text).

The level width associated with the outer part of the QD in the absence of the 
inner puddle can be identified by  
expanding~(\ref{eq:QD_transmission_amplitude_sup}) near a resonance, and singling 
out the quantity that plays the role of the Lorentzian width.  Assuming 
$r_{\mathrm{B}} r'_{\mathrm{A}}$ to be real, resonance transmission occurs at 
integral multiples of $\epsilon = 2\pi$.  Expansion around $\epsilon = 0$ 
gives $\gamma = \left( 1 - r'_{\mathrm{A}} r_{\mathrm{B}} \right) / 
r'_{\mathrm{A}} r_{\mathrm{B}}$. Assuming that the two tunnel-bridges that 
define the QD have equal transmission and reflection probabilities, and 
substituting the foregoing expression in~(\ref{eq:QD_transmission_amplitude_sup}), 
one obtains Eq.~(2) of the main text up to an unimportant constant phase 
factor.

% {\center \textbf{Initial wave function of a localized level tunnel-coupled 
% to lead} }

\subsection*{Initial wave function of a localized level tunnel-coupled to lead 
}

At zero temperature the isolated subsystem composed of the localized state and 
the tunnel-coupled lead is a many-body system, whose wave function can be 
written as a linear combination of Slater determinants in the basis states 
$\left\{ \sigma;
n_{k_{1}}, n_{k_{2}}, \ldots, n_{k_{N}} \right\} \equiv \left\{ \sigma; \xi 
\right\}$. Here $\sigma = 0,1$ denotes an empty or occupied localized state
and $n_{k_i}$ is the occupation of the state $k_i$ in the lead (the lead 
accommodates $N$ single-particle states). Formally,
\begin{multline}
  \left|\mathrm{GS}\right\rangle =\sum_{\sigma;\xi}c_{\sigma;\xi}\left|\left\{ 
  \sigma;\xi\right\} \right\rangle \\
  =\sum_{\xi}c_{0;\xi}\left|\left\{\sigma = 0;\xi\right\} \right\rangle 
  +\sum_{\xi}c_{1;\xi}\left|\left\{\sigma = 1;\xi\right\} \right\rangle \\
  =\left|\sigma = 0\right\rangle \otimes\left|\mathrm{FS},0\right\rangle 
  +\left|\sigma = 1\right\rangle \otimes\left|\mathrm{FS},1\right\rangle \,.
  \label{eq:GS}
\end{multline}
Here $\left|\left\{ \sigma;\xi\right\} \right\rangle$ is a Slater determinant 
built from the corresponding occupied states and $c_{\sigma;\xi}$ is the 
associated amplitude. The ket $\left|\mathrm{FS},0\right\rangle 
\equiv\sum_{\xi}c_{0;\xi}\left|\left\{ 0;\xi\right\} \right\rangle$ physically 
denotes the Fermi sea when the localized state is empty and similarly for 
$\left|\mathrm{FS},1\right\rangle$.

Thus, it is found that the environment can be treated as a two-state system 
with $\left|0\right\rangle \equiv\left|\sigma=0\right\rangle 
\otimes\left|\mathrm{FS},0\right\rangle $
and $\left|1\right\rangle \equiv\left|\sigma=1\right\rangle 
\otimes\left|\mathrm{FS},1\right\rangle$.

% {\center \textbf{Mean occupation of localized level} }
\subsection*{Mean occupation of localized level }

The mean occupation of the localized state can be calculated as follows. The 
retarded Green's function of the localized level, which is coupled to outer 
leads, is
$ G^{\mathrm{R}}(E) = (E - \epsilon_{0} + i\Gamma)^{-1}$. Here $\epsilon_0$ is 
the level's eigenenergy and $\Gamma$ is the level's width.  This yields a 
spectral density $B(E) = -2 \Im \mathrm{Tr} G^{\mathrm{R}} (E) = 
2 \Gamma / [(E - \epsilon_{0})^{2} + \Gamma^{2}]$.
The mean occupation is then given by
\begin{equation}
n = \int_{-\infty}^{\infty}\frac{\mathrm{d}E}{2\pi}\frac{1} {e^{(E-\mu)/T}+1} 
\frac{2\Gamma}{(E-\epsilon_{0})^{2}+\Gamma^{2}}\,.
\label{eq:red_dot_occupation_probability}
\end{equation}
Calculation of the integral in the limit of $T \gg \Gamma$ gives that the mean 
occupation equals the Fermi-Dirac function, $n = [e^{(\epsilon_0 - \mu) / T} + 
1]^{-1}$.
In the opposite limit of zero temperature the Fermi-Dirac function in the 
integrand of Eq.~(\ref{eq:red_dot_occupation_probability}) is the step 
function
$\Xi(\mu - E)$, where $\Xi(x) = 1$ ($x>0$) or $\Xi(x) = 0$ ($x<0$).  
Eq.~(\ref{eq:red_dot_occupation_probability}) then gives
\begin{equation}
n = \frac{1}{\pi}\left[\arctan\left(\frac{\mu-\epsilon_{0}}{\Gamma}\right)
+\frac{\pi}{2}\right] \,.
\label{eq:red_dot_occupation_probability_T=0_result_}
\end{equation}

% {\center \textbf{The density matrix of the composite system} }
\subsection*{The density matrix of the composite system }

We are interested in the wave function of the whole system that corresponds to 
the interfering electron being either scattered by the QD or transmitted 
through the lower Mach-Zehnder interferometer (MZI) arm. This wave function 
can be written as
\begin{equation}
  \left| \psi \right\rangle = \left| \psi_{\mathrm{u}} \right\rangle + \left| 
  \psi_{\mathrm{d}} \right\rangle
  \label{eq:psi}
\end{equation}
with
\begin{eqnarray}
  \left| \psi_{\mathrm{u}} \right\rangle  & = & \frac{1}{\sqrt{2}} \left[ 
    t_{\mathrm{QD}}(1) \left| \mathrm{u} \right\rangle \otimes\sqrt{n} \left| 
    1 \right\rangle +t_{\mathrm{QD}}(0) \left| \mathrm{u} \right\rangle
    \otimes \sqrt{1-n} \left| 0 \right\rangle \right. \nonumber \\
    &  & \left. + r_{\mathrm{QD}}(1) \left| \mathrm{w} \right\rangle 
    \otimes\sqrt{n}\left|1\right\rangle +r_{\mathrm{QD}}(0)\left| \mathrm{w} 
    \right\rangle \otimes \sqrt{1-n} \left| 0 \right\rangle \right] \,,
   \nonumber \\
  \left|\psi_{\mathrm{d}}\right\rangle  & = & 
  \frac{1}{\sqrt{2}}e^{i\phi}\left| \mathrm{d} \right\rangle 
  \otimes\left(\sqrt{1-n}\left|0\right\rangle +\sqrt{n}\left|1\right\rangle 
  \right)\,.
  \label{eq:psi_d_psi_u}
\end{eqnarray}
Here $\left| \mathrm{d} \right\rangle$ is the lower partial wave that goes 
directly towards the second junction,  and interferes with the part of the 
upper partial wave that is transmitted through the QD, $\left| \mathrm{u} 
\right\rangle$. The other part of the upper partial wave, namely the one which 
is reflected from the QD to another drain, is denoted by $\left| \mathrm{w} 
\right\rangle$. The phase $\phi$ is
defined in the main text after Eq.~(3). The transmission amplitudes 
$t_{\mathrm{QD}}(1)$ and $t_{\mathrm{QD}}(0)$ are abbreviations for 
$t_{\mathrm{QD}}(\epsilon,\theta,\theta_{\mathrm{R}})$ and 
$t_{\mathrm{QD}}(\epsilon,0,0)$, respectively (cf.  Eq.~(2) in the main text).  
The reflection amplitudes $r_{\mathrm{QD}}(1)$ and $r_{\mathrm{QD}}(0)$ are 
similarly defined.

The density matrix used in Eq.~(3) of the main text is then $\hat{\rho} = 
\left| \psi \right\rangle \left\langle \psi \right|$.

% {\center \textbf{The projection operator $\hat{D}$} }
\subsection*{The projection operator $\hat{D}$ }

The operator $\hat{D}$ is essentially a projection operator, which selects 
only the part of the wave function that arrives at one drain of the MZI.
It can be obtained as follows. The two junctions of the MZI define three 
regions, each of which consists of two segments that accommodate the 
propagation of a partial wave.  Formally, the state of an electron in the 
region after the second junction can be represented by a two-component wave 
function with basis vectors $| \mathrm{u'} \rangle$ and $| \mathrm{d'} 
\rangle$, corresponding to a propagation towards the upper or lower drain, 
respectively.  In this basis the operator $\hat{D}' = | \mathrm{u'} \rangle 
\langle \mathrm{u'} |$, assuming the upper drain is the one which is measured 
in the experiment.  In the main text, though, we use the basis vectors $| 
\mathrm{u} \rangle$ and $| \mathrm{d} \rangle$, which correspond to the two 
segments of the MZI in the regions just in front of the second junction. Hence 
we need to perform the transformation $\hat{D} = \hat{S}\da \hat{D}' \hat{S}$, 
where $\hat{S}$ is the scattering matrix of the second junction. A valid 
scattering matrix describing a symmetric junction is
\begin{equation}
  \hat{S} = \frac{1}{\sqrt{2}} \left(
  \begin{array}{cc}
   \phantom{-} 1 & 1\\
            -  1 & 1
  \end{array}\right)\,.
  \label{eq:OP_basis_2}
\end{equation}
Using that matrix one obtains that $\hat{D}$ is a $2 \times 2$ matrix with all 
entries equal to $1/2$. Since neither the arrival to the drain nor the passage 
through the second junction alters the state of the localized level, one 
arrives at the definition
$\left(\left\langle \sigma\right|\otimes\left\langle 
s\right|\right)\hat{D}\left(\left|s'\right\rangle 
\otimes\left|\sigma'\right\rangle \right)=\delta_{\sigma\sigma'}/2$
for all combinations of $s, s' = \mathrm{u}, \mathrm{d}$.

% {\center \textbf{Charge stability diagram and scenarios for changes of 
% occupancy }}

\subsection*{Charge stability diagram and scenarios for changes of occupancy }

\begin{figure}[tbp]
\begin{center}
% \vspace{0cm}
\includegraphics[width=0.49\textwidth]{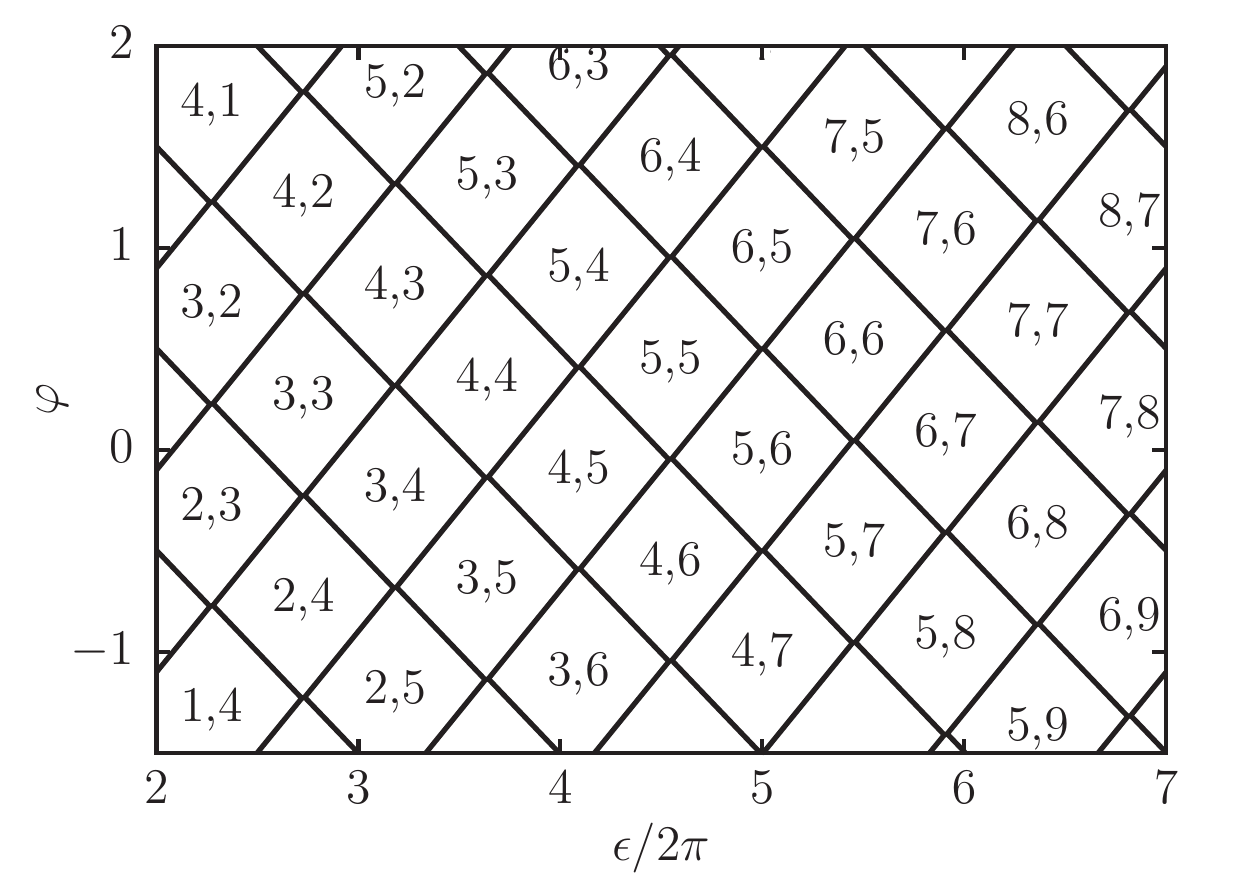}
% \vspace{-0.5cm}
\caption[] {\label{fig:stability_diagram_W}
Example of a stability diagram describing the charge distribution in the QD in 
the weak coupling regime.  The pair of numbers in each polygon denotes the 
equilibrium occupation of the outer and inner regions, $\No, \Ni$.  Here we 
set $\Ko =  1$, $\Ki = 1.5$, $\Koi = 0.02$ and $(\beta / \alpha) \cdot (\Delta 
/ \Di) = 1.2$.
} % caption
\end{center}
% \vspace{-0.5cm}
\end{figure}

\begin{figure}[tbp]
\begin{center}
% \vspace{0cm}
\includegraphics[width=0.49\textwidth]{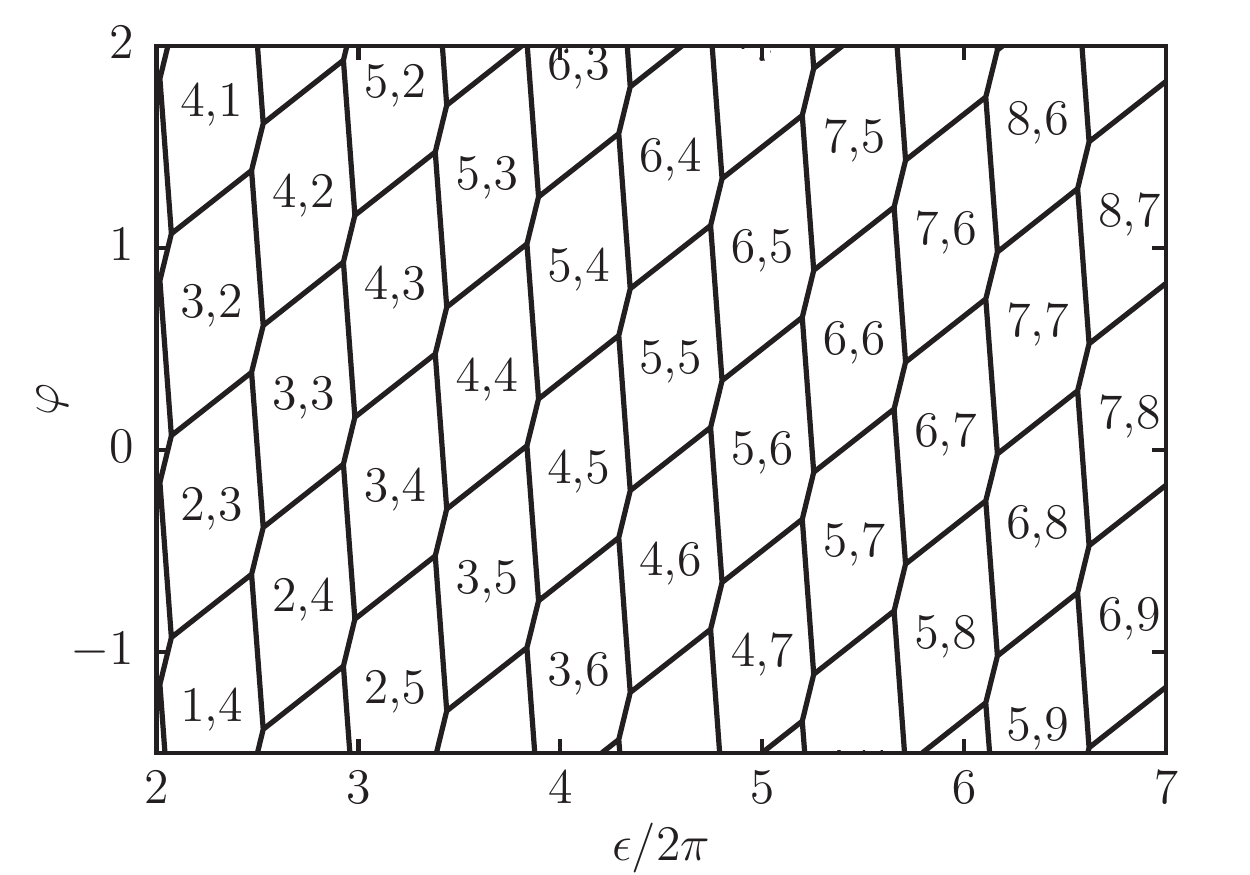}
% \vspace{-0.5cm}
\caption[] {\label{fig:stability_diagram_S}
Example of a stability diagram describing the charge distribution in the QD in 
the strong coupling regime. Here we set $\Ko =  1$, $\Ki = 1.5$, $\Koi = 
0.85$ and $(\beta / \alpha) \cdot (\Delta / \Di) = 1.2$.
} % caption
\end{center}
% \vspace{-0.5cm}
\end{figure}

We model the outer and inner parts of the QD by conductors, and write the 
total energy of the QD as
\begin{multline}
  E = \frac{\Ko}{2} \left(\No - \frac{\epsilon}{2\pi} - \varphi \right)^{2} + 
  \frac{\Ki}{2} \left(\Ni - \frac{\beta}{\alpha} \frac{\Delta}{\Di} 
  \frac{\epsilon}{2\pi} + \varphi \right)^{2} \\
  + \Koi \left(\No - \frac{\epsilon}{2\pi} - \varphi \right) \left(\Ni - 
  \frac{\beta}{\alpha} \frac{\Delta}{\Di} \frac{\epsilon}{2\pi} + \varphi 
  \right) \,.
  \label{eq:E_function}
\end{multline}
Here $\Ko$, $\Ki$ and $\Koi$ are positive and can be related to the conductors 
capacitance matrix. The number of flux quanta penetrating the QD is $\varphi$.  
The charges in the outer and inner regions, $\No$ and $\Ni$, respectively, are 
taken to be integers. They are determined by the requirement that the energy 
function be minimal for given values of all the other parameters. The 
parameters $\Delta$ and $\Di$  are the level spacing of the outer and inner 
regions, respectively (the former is defined in the main text).  The parameter 
$\epsilon$ is the dimensionless gate voltage appearing in the main text and is 
related to the gate voltage $V_g$.  It enters the energy function in such a 
way that --- in the absence of interactions between the inner and the outer 
regions ---  $\No$ ($\Ni$) increases by one when $\alpha V_g$ ($\beta V_g$) 
increases by $\Delta$ ($\Di$).  Figs.~\ref{fig:stability_diagram_W} 
and~\ref{fig:stability_diagram_S} show examples of stability diagrams.

% === Figure : stability diagram weak>>>

As $V_g$ is swept, several scenarios (charge variations) may occur. The 
possible scenarios and their probabilities can be inferred from the stability 
diagram. This is simply the relative length of the appropriate
polygon edge projected on the $\varphi$-axis. Note that only 
3 independent parameters determine the probability of
the scenarios; they can be chosen as $\Ko / \Koi$, $\Ki / \Koi$ and $(\beta
/ \alpha) \cdot (\Delta / \Di)$.

The relation between the energy function and the transmission phase is 
established in the following way.  We first show that the screening is given 
by $\theta = 2 \pi \Koi / \Ko$.  This can be obtained by treating $\No$ and 
$\Ni$ as continuous variables for a moment, and requiring that $\partial E / 
\partial \No = 0$.  One then obtains that, if $\Ni$ increases by 1, then $\No$ 
varies by $-\Koi/\Ko$. In accordance with Eq.~(2) of the main text we identify 
that $\theta = 2 \pi \Koi / \Ko$.
The parameter $c$ in the main text is chosen so as to fit the investigated 
scenario. For instance, to investigate population switching we tune $c$ such 
that the inner puddle has mean occupation $1/2$ when the outer region is in a 
valley and $\theta$ is sufficiently large to shift the outer region beyond a 
peak.

The specific values of the parameters appearing in~(\ref{eq:E_function}) 
determine the stability diagram of the QD. These are sample-dependent, and can 
be estimated for a given experimental realization. To exemplify the type of 
analysis one may perform, and to gain some insight into the physics, we 
concentrate on two extreme regimes, namely the regime where the interaction 
between the inner and outer regions of the QD is weak (``W''), and the regime 
where the interaction is strong (``S''). For concreteness, we take $(\beta / 
\alpha) \cdot (\Delta / \Di) = 1.2$.
Moreover, we assume that the inner puddle has a slightly higher charging 
energy than the outer region, and put $\Ko = 1$ and $\Ki = 1.5$.

% === Figure : stability diagram strong>>>

\emph{Weak coupling regime} --- Taking $\Koi = 0.02$, which implies $\theta = 
0.04 \pi$, yields that the probable scenarios are $(\No, \Ni) \rightarrow
(\No, \Ni + 1)$ (W(a)) and $(\No, \Ni) \rightarrow (\No + 1, \Ni)$;  
population switching $(\No, \Ni) \rightarrow (\No - 1, \Ni + 1)$ (W(b)) is 
rare (cf. Fig.~\ref{fig:stability_diagram_W}). In W(a) the outer region is not 
too close to a resonance, and $\zeta\simeq e^{i\theta_{\mathrm{R}}}$. Since 
$\theta$ is small, presumably also $\theta_{\mathrm{R}} \simeq 0$, namely the 
screening takes place somewhere else --- neither in the outer region of the 
dot nor in the upper MZI arm. Then one may observe a sharp signature in the 
transmission phase, but not a discontinuity. We analyse W(b) by writing 
$\theta = m\gamma$ and $\epsilon = p\gamma$ where $m$ and $p$ are positive 
parameters that fulfill $m > p$.  Expansion of $\zeta$ around small values of 
$m \gamma$ and $p \gamma$ yields $\zeta \simeq \left(1 - ip \right) 
e^{i\theta_{\mathrm{R}}} / \left( 1 - ip + im \right)$.  Then for $m > p + 
1/p$ a phase lapse occurs if $\theta_{\mathrm{R}} = - \arctan \left[ m / 
  \left[ p \left(m - p
\right) - 1 \right] \right]$. In a generic case where $m \ll p(m-p) - 1$ a 
phase lapse occurs if $\theta_{\mathrm{R}} \simeq 0, 2\pi$. 

\emph{Strong coupling regime} --- Taking $\Koi = 0.85$, which implies $\theta 
= 1.7 \pi$, yields that the probable scenarios are $(\No, \Ni) \rightarrow 
(\No + 1, \Ni)$, the scenario $(\No, \Ni) \rightarrow (\No, \Ni + 1)$ (S(a)) 
and the scenario $(\No, \Ni) \rightarrow (\No - 1, \Ni + 1)$ (S(b)) (cf.  
Fig.~\ref{fig:stability_diagram_S}). We analyse S(a) by writing $\theta = 2\pi 
- m\gamma$ and $\epsilon = 2\pi - p\gamma$ where $m$ and $p$ are positive 
parameters that fulfill $m > p$.  Expansion of $\zeta$ around small values of 
$m \gamma$ and $p \gamma$ yields $\zeta \simeq \left(1 + ip \right) 
e^{i\theta_{\mathrm{R}}} / \left( 1 + ip -im \right)$.  Then for $m > p + 1/p$ 
a phase lapse occurs if $\theta_{\mathrm{R}} = \arctan \left\{ m / \left[
p \left(m - p \right) -1 \right] \right\}$. In a generic case where $m \ll 
p(m-p) - 1$ a phase lapse occurs if $\theta_{\mathrm{R}} \simeq 0, 2\pi$. In 
S(b) one has $\zeta \simeq e^{i \theta_{\mathrm{R}}}$, namely $\left \langle 
\zeta \right \rangle \simeq 1 - n + n e^{i \theta_{\mathrm{R}}}$. This gives a 
phase lapse for $\theta_{\mathrm{R}} \simeq \pi$.

\end{document}